# Ultrafast and Nanoscale Plasmonic Phenomena in Exfoliated Graphene Revealed by Infrared Pump-Probe Nanoscopy

*Martin Wagner[†], Zhe Fei[†], Alexander S. McLeod[†], Aleksandr S. Rodin[†,I], Wenzhong Bao[‡,$], Eric G. Iwinski[†], Zeng Zhao[$], Michael Goldflam[†], Mengkun Liu[†], Gerardo Dominguez[∥,⊥], Mark Thiemens[⊥], Michael M. Fogler[†], Antonio H. Castro Neto[□,I], Chun Ning Lau[$], Sergiu Amarie[■], Fritz Keilmann[=], and D. N. Basov[\*,†]*

[†] Department of Physics, University of California, San Diego, La Jolla, California 92093, USA

[I] Department of Physics, Boston University, 590 Commonwealth Avenue, Boston MA 02215, USA

[‡] Materials Research Science and Engineering Center, University of Maryland, College Park, Maryland 20742, USA

[$] Department of Physics and Astronomy, University of California, Riverside, California 92521, USA

[∥] Department of Physics, California State University, San Marcos, San Marcos, California, 92096, USA

[⊥] Department of Chemistry and Biochemistry, University of California, San Diego, La Jolla, California 92093, USA

[□] Graphene Research Centre and Department of Physics, National University of Singapore, 117542, Singapore

[■] Neaspec GmbH, Bunsenstr. 5, 82152 Martinsried, München, Germany

[=] Ludwig-Maximilians-Universität and Center for Nanoscience, 80539 München, Germany

KEYWORDS: Graphene, Dirac plasmon, pump-probe, time-resolved spectroscopy, near-field microscopy, s-SNOM

ABSTRACT

Pump-probe spectroscopy is central for exploring ultrafast dynamics of fundamental excitations, collective modes and energy transfer processes. Typically carried out using conventional diffraction-limited optics, pump-probe experiments inherently average over local chemical, compositional, and electronic inhomogeneities. Here we circumvent this deficiency and introduce pump-probe infrared spectroscopy with ~20 nm spatial resolution, far below the diffraction limit, which is accomplished using a scattering scanning near-field optical microscope (s-SNOM). This technique allows us to investigate exfoliated graphene single-layers on $SiO_2$ at technologically significant mid-infrared (MIR) frequencies where the local optical conductivity becomes experimentally accessible through the excitation of surface plasmons via the s-SNOM tip. Optical pumping at near-infrared (NIR) frequencies prompts distinct changes in the plasmonic behavior on 200 femtosecond (fs) time scales. The origin of the pump-induced, enhanced plasmonic response is identified as an increase in the effective electron temperature up to several thousand Kelvin, as deduced directly from the Drude weight associated with the plasmonic resonances.

Graphene is emerging as a promising system for optoelectronic and plasmonic applications.[1-4] This is in part because the optical, electronic and plasmonic phenomena present in graphene are amenable to modifications by chemical doping and electrostatic gating. Specifically, these stimuli influence both the amplitude and the wavelength of Dirac plasmons, the surface charge oscillations in graphene.[5,6] Tunable surface plasmons are of high interest in the context of confining and controlling long-wavelength electromagnetic radiation at the length scales typical of x-rays.[7] The key novelty of the results presented here is that the plasmonic effects in graphene can be modified on ultrafast time scales with an efficiency rivaling that of electrostatic gating. This is achieved using a novel approach that enables pump-probe spectroscopy in the infrared (IR) region with nanoscale spatial resolution. Common optical pump-probe experiments have been extensively employed to investigate fundamental physical effects in condensed matter.[8,9]

However, the spatial resolution of these techniques is typically limited by diffraction restricting the minimum sampling area to about 1 $\mu m^2$ for measurements in the visible and to $10^2$-$10^4$ $\mu m^2$ or more for IR and THz studies. For that reason, pump-probe experiments on graphene have been performed on large area samples[10-13] that are prone to the formation of various defects or grain boundaries[14,15] obscuring intrinsic electronic and optical phenomena. These irregularities are absent in exfoliated graphene with typical dimensions smaller than the wavelength of IR light: the type of specimens we investigated in our experiments.

Here, we implemented a nano-spectroscopic IR local probe via s-SNOM[16,17] under intense NIR laser excitation. We employed this new technique to explore the ultrafast response of Dirac plasmons in graphene. Our key observation pertains to strong enhancement of plasmonic features by NIR pumping, which we directly monitor in our time-resolved spectra. We can fully account for this behavior by modeling NIR pumping as light-induced heating of the free-carrier subsystem of graphene. Specifically, we show that the intraband Drude weight $D$ due to Dirac quasiparticles increases with their effective carrier temperature even though the density of electrons in the system, i.e. the number of electrons in the conduction band minus the number of holes in the valence band, remains unchanged under NIR illumination. We stress that the Drude weight is directly accessible in our measurements through the analysis of the plasmonic features in the infrared spectra that we collect with the 200 fs temporal resolution. Furthermore, our data attest to the ability to optically modify Dirac plasmons on 20 nm length scales and 200 fs time scales. An important virtue of our approach is that the conversion of light to electron-hole pairs can now be investigated at length scales characteristic of these light-matter interaction processes and of prototypical photonics/plasmonics devices.[1,2,4,7,18,19]

Our setup (Figure 1a) integrates a commercial s-SNOM (Neaspec GmbH) with an MIR source (Lasnix) emitting ultrashort (<100 fs) pulses based on a 40 MHz NIR fiber laser (Toptica Photonics, Inc.).[20-22] We note that owing to field enhancement by the s-SNOM tip[23] an in-plane MIR peak field strength of the order of $10^2$ kV/cm can be obtained with such a relatively low-cost, high repetition rate laser system (see Supporting Information (SI)). Ultrafast temporal resolution is achieved by operating the s-SNOM in pump-probe mode whereby the synchronized 100 fs NIR pulses of the fiber laser serve as 1.56 μm pump light with variable delay. Exfoliated graphene samples on 300 nm $SiO_2$ were studied at ambient conditions. Raman spectroscopy confirmed an unintentional hole-doping $|n| = (2.6 \pm 1) \times 10^{12}$ cm$^{-2}$, corresponding to a chemical potential $\mu_0 = -(1500 \pm 300)$ cm$^{-1}$.

The near-field response of graphene on SiO$_2$ in the MIR region is dominated by hybrid resonances originating from the coupling of plasmons to the surface phonon modes of the substrate.[24] We first consider these plasmonic features in the near-field data in the absence of the NIR excitation. In Figure 2a we plot $s(\omega)$ spectra: the second harmonic of the near field scattering amplitude $s_2$ normalized to that of silicon. The $s(\omega)$ spectrum of SiO$_2$ alone (black curve in Figure 2a) is dominated by two surface phonons denoted as α and β, giving rise to a strong peak at $\omega_\alpha = 1125$ cm$^{-1}$ and a weaker one at $\omega_\beta = 785$ cm$^{-1}$.[25] A single layer of graphene modifies the main features of the SiO$_2$ substrate response (red curve). We register a 5 cm$^{-1}$ blueshift of the α resonance compared to SiO$_2$ and a broadening of its high-frequency shoulder, effects rooted in the interaction of the surface phonons of the SiO$_2$ substrate with the Dirac plasmons in graphene.[24] Note that the plasmon-phonon coupling is fundamentally a finite momentum effect (see dispersion plot in Figure 3a). The ability of near-field nanoscopy to investigate this effect stems from the AFM tip providing access to momenta of the order of $q_{\text{tip}} \sim 1/a$ (vertical dashed lines in Figure 3a,b) where $a$ is the tip radius.[24] Thus we are able to study surface plasmons in pristine, exfoliated graphene. We note that plasmonic coupling in far-field experiments requires nano-structuring of large-area CVD or epitaxial samples[26] prone to defects inherent to both growth[14,15] and patterning[26] procedures.

Because our inquiry into the ultrafast plasmon dynamics relies on monitoring the hybrid plasmon-phonon resonances, we discuss the dispersion relationships unraveling the physics of this coupling to set the stage for later analysis of the pump-probe data. The dispersion of the Dirac plasmon alone (red curves in Figure 3a,b) and of the hybrid plasmon-phonon modes (color plots in Figure 3a,b) is calculated[24] based on the standard sheet conductivity of graphene comprising intra- and inter-band components as:

$$\sigma(\omega) = \sigma_{\text{intra}}(\omega) + \sigma_{\text{inter}}(\omega), \qquad \sigma_{\text{intra}}(\omega) = \frac{i}{\pi}\frac{D}{\omega + i\gamma}. \qquad (1)$$

The plasmonic response of graphene is dominated by the intraband part that is defined by two parameters: the Drude weight $D$ and the scattering rate $\gamma$.[27] Ignoring for a moment the impact of the substrate, the dispersion of the plasmon (red curves in Figure 3a,b) follows the familiar square-root behavior $\omega \sim (k_F q)^{1/2}$ for $q$ values below the Fermi wavevector $k_F$. The horizontal lines in Figure 3a,b denote the SiO$_2$ surface phonons, defined by Re $\kappa(\omega) = 0$, where $\kappa(\omega) = \frac{1+\varepsilon_{SiO2}(\omega)}{2}$ is the arithmetic mean of the dielectric functions of air and SiO$_2$. Once coupling between the plasmon and SiO$_2$ phonons is taken into account the dispersion is dominated by avoided crossings. An immediate consequence of the coupling is the blueshift of the α resonance that is consistent with the data in Figure 2a.

Next, we examine the influence of NIR pumping on the graphene response at zero time delay between pump and probe pulses. We present pump-induced changes of the scattering amplitude Δ$s$(ω)/$s$(ω) (Figure 2b), collected from a ~10-20 nm spot that is defined by the radius of the tip. The spectra show an increase of the scattering amplitude, which is particularly pronounced at frequencies just above the α and β modes, amounting to as much as 20% around $\omega =$ 1200 cm$^{-1}$. The upper frequency cut-off in Figure 2b is governed by the signal-to-noise of our data. At $\omega = 900$ cm$^{-1}$ a weaker and broader peak appears whereas the β mode is only marginally amplified. Multilayer graphene shows a stronger photoinduced response (see inset of Figure 2b) that grows nearly linearly with the number of layers.

Diffraction-limited pump-probe experiments on graphene have established several facts relevant to our nanoscale observations.[12,13,28,29] Immediately after photoexcitation graphene reveals distinct electron and hole chemical potentials that last only 10-30 fs after the pump pulse arrives. Interband scattering of electrons and holes occurs via Auger and carrier multiplication processes that are kinematically allowed and efficient in graphene. The net effect is that 30 fs after the pump pulse is incident on graphene, electrons and holes are characterized by a single Fermi-Dirac distribution with the same *effective* temperature $T$ and chemical potential $\mu(T)$.[12,28] The 200 fs temporal resolution of our apparatus only enables access to the latter regime of "hot electrons" whereas the effects of rapid equilibration are beyond our range.

Remarkably, the spectral changes associated with pump-induced heating of electrons in single-layer graphene closely resemble those originating from electrostatic gating. As shown in Figure 2b (dots), similar Δ$s$(ω)/$s$(ω) spectra are obtained by applying a gate voltage of V$_g$ = -70V and thus increasing the sample's hole-doping. To the best of our knowledge, data in Figure 2b is the first attempt to explicitly compare spectral features induced in graphene by gating and ultrafast photoexcitation. We were able to account for this peculiar similarity of the response to the two distinct stimuli within our intraband Drude model of plasmonic response introduced in the following section.

We now proceed to our quantitative description of the plasmonic response in the regime of pump-induced heating of the electronic system based on eq 1 and introduce a temperature dependent intraband Drude weight (Figure 3c) in the following form[30]:

$$D = \left(\frac{2e^2}{\hbar^2}\right) k_B T \ln\left[2 \cosh\left(\frac{\mu}{2k_B T}\right)\right]. \qquad (2)$$

In the zero-temperature limit eq 2 reduces to $D = \left(\frac{e^2}{\hbar^2}\right)\mu \propto \sqrt{|n|}$ where $n$ is the carrier density[27], i.e. the electron concentration in the conduction band *minus* the hole concentration in the valence band. At elevated temperatures, $k_B T \gg \mu$, eq 2 implies a linear dependence $D \propto T$ (asymptote in Figure 3c). Counterintuitively, for $T \neq 0$ a peculiar situation can arise: $D$ increases with temperature although the density of electrons in the system $n$ is constant[31], a scenario relevant to the photoinduced heating of our graphene samples. The net effect of high electron temperature is that the plasmonic response of graphene at IR frequencies is enhanced as we will show below.

The Drude model of plasmonic response grasps the gross features of both the transient and the equilibrium data in Figure 2. At equilibrium (Figure 2a), this model accounts for the spectral features associated with both the α and β plasmon-phonon modes assuming $\gamma = 100$ cm$^{-1}$ and the Drude weight $D_0 = 24.3|\mu_0|\sigma_0/\hbar$ (see Figure S2 in SI). Here $\sigma_0 = e^2/(4\hbar)$ denotes the so-called universal conductivity of graphene. Owing to small signal-to-noise ratios the full MIR intensity with an in-plane field strength of the order of $10^2$ kV/cm is necessary to carry out the pump-probe experiment (see SI). This results in an increase of the Drude weight up to $D_{MIR} = 1.10D_0$ which can be attributed to the elevated electron temperature $T = 1500$ K according to eq 2.[28,32] Nevertheless, in the absence of NIR pumping, the dominant effect in the spectra (see Fig. S2 in the SI) is due to the three-fold increase of the scattering rate to $\gamma = 300$ cm$^{-1}$.[33] An increase in carrier temperature dominates in our NIR pumping results. Specifically, we were able to reproduce the pump-induced enhancement of the near-field amplitude at $\omega > 1050$ cm$^{-1}$ with the following values for $D$: $D_{2mW} = 1.21D_0$, $D_{5mW} = 1.33D_0$ and $D_{10mW} = 1.46D_0$ (blue, red and black line in Figure 2c). The full set of modeling parameters is presented in Table 1. We found that the model spectra capture the essential aspects of our data even if we keep the scattering rate constant at $\gamma = 300$ cm$^{-1}$ for any NIR illumination power. At frequencies below 1000 cm$^{-1}$, there is a slight mismatch between the details of the experimental and theoretical spectra. The simulation correctly replicates a peak in the vicinity of the β mode. However, the peak width is smaller as compared to experimental data. The net result of these simulations is that hot electrons in graphene (with an effective temperature in excess of 2000 K) lead to significant enhancement of the plasmonic resonance. Electrostatic gating is replicated in our modeling by tuning the Fermi energy at constant temperature ($T = 1500$ K) and scattering rate ($\gamma = 300$ cm$^{-1}$), yielding the Drude weight of $1.46 D_0$, similar to the case of 10 mW optical excitation. The corresponding gate-induced carrier density of $|\Delta n| = 4.2 \times 10^{12}$ cm$^{-2}$ is close to the experimental value of $5 \times 10^{12}$ cm$^{-2}$ estimated from the common capacitor model for the applied gate voltage $V_g = -70$V.[34] However, under ambient conditions the charge-neutrality point is known to shift in electrostatic gating.[35] Consequently, the actual carrier density in our gating experiment can be expected to differ from the value inferred from the capacitor model, especially in the unexplored regime of additional, intense MIR illumination.

It is instructive to discuss the response of hot electrons and plasmons in graphene under NIR pumping in terms of the plasmon-phonon dispersion relations (Figure 3a,b). In Figure 3a we plot the dispersion under equilibrium conditions corresponding to the intraband Drude weight $D_0$. The dispersion in Figure 3b represents the pump-induced non-equilibrium situation and is different from the dispersion in Figure 3a in two ways. First, the dispersion traces are broadened by excessive damping ($\gamma = 300$ cm$^{-1}$) caused by the hot electron temperature as in the case of pure MIR pumping at full intensity (see SI). Second, the Drude weight governing the behavior of the plasmonic branch is increased by nearly 50% compared to the equilibrium value. The elevated Drude weight increases the slope of the plasmon dispersion (red line). Enhanced overlap with the SiO$_2$ phonon modes leads to a blue shift of both the α and β resonances at $q \sim q_{tip}$. Thus, spectral changes induced by NIR pumping indeed are expected to be similar to those originating from electrostatic gating despite the assumption of constant electron density in our modeling. Note that intense in-plane DC fields have recently been predicted to increase the slope of the plasmon dispersion.[36] However, this is a higher-order effect that is not expected to play a significant role in our case. That is because in our experiments the AC electric fields average to zero both in time, and, due to the radial field distribution directly under the tip, also in space.

We stress that even though $\mu(T)$ is suppressed at elevated temperatures (Figure 3c), the Drude weight $D$ is enhanced above its equilibrium value $D_0$ for $T > 0.60 \, |\mu_0|/k_B = 1270 \, K$ (black square). The sensitivity of the Drude weight to electronic temperature enables ultrafast optical control of plasmonic effects in graphene at fs time scales. Since $D$ defines both the amplitude and wavelength of surface plasmons in graphene, both of these fundamental characteristics will be impacted by photoexcitation.[5,6]

For bi- and trilayer graphene, similar behavior showing enhanced intraband Drude weight with temperature is expected when the electron temperature $k_B T$ approaches (or exceeds) the interlayer tunneling energy $\gamma_1 = 0.4$ eV, a situation close to our experimental parameters. An approximately linear dispersion in that region should mimic the temperature dependence of $D$ in single-layer graphene. Data in the inset of Figure 2b and comparable time constants in Figure 4 qualitatively support this interpretation, but quantitative comparison warrants further study.

We now analyze the temporal evolution of the near-field plasmonic response by measuring the spectrally integrated scattering amplitude. In these latter experiments, the probe beam was centered at a frequency close to 1200 cm$^{-1}$ (see Figure 1b). Thus the scattering amplitude is dominated by the high-frequency flank of the α plasmon-phonon mode where pump-induced changes are the most distinct, according to Figure 2b. We have carried out these experiments for a sample with terraces of single-, bi- and tri-layer graphene (Figure 4c,d). Our assignment of the number of layers in each terrace is based on optical contrast and is confirmed by Raman micro-probe spectroscopy. In Figure 4a we plot the photo-induced changes of the scattering signal Δ$s$/$s$ for various pump-probe time delays Δ$t$ collected from an area as small as ~500 nm$^2$. We found that the temporal traces show a biexponential time dependence[10,37] and extracted the two time constants (Figure 4b). Time-resolved imaging data are presented in Figure 4e. Here we display images of NIR photo-induced changes Δ$s$/$s$ for different pump-probe time delays. A significant increase of the scattering amplitude occurs at temporal overlap (Δ$t$ = 0). Naturally, the signal in the region of the SiO$_2$ substrate is not impacted by the NIR pump. Few-layer graphene exhibits stronger pump-induced changes in scattering amplitude compared to single-layer terraces. This is expected since the absorption cross-section of the pump light scales with the number of layers. For increasing time delay Δ$t$, the sequence in Figure 4e displays a decrease in the pump-induced Δ$s$/$s$ signal. Note that our time-resolved imaging does not reveal significant spatial variations in the observed decay constants.

Below, we briefly outline the key features revealed by the temporal profiles of our pump-probe data. The rise time in Figure 4a is limited by the temporal resolution of the MIR probe pulses (200 fs; see SI). The positive sign of the pump-induced signal is consistent with the idea of increased intraband Drude weight $D$ that we have established based on the analysis of the pump-probe spectra in Figure 2b. We assign the faster time scale $\tau_1 \approx 200$ fs to cooling of the hot carriers via optical phonon emission.[12,28,37] Analysis of the spectra informed us that this cooling process occurs in the regime of constant electron density even though the Drude weight is suppressed with decreasing carrier temperature. The faster time constant is quite similar for the heavily doped single- and few-layer graphene. The longer time constant $\tau_2$ is assigned to energy relaxation involving acoustic phonons.[12,37] Bi- and tri-layer samples show larger $\tau_2$ values: a plausible consequence of energy diffusion to the substrate.[37]

In conclusion, we reported near-field pump-probe spectroscopy based on s-SNOM combining exceptional spatial, spectral and temporal resolution. Experimental capabilities of our ultra-fast IR nanoscope significantly extend previously reported results[20,38-42]. Well-established fiber-based near-field methods[43,44] may be easier to implement but are unfortunately limited to visible and NIR frequencies. In contrast, ultra-fast s-SNOM described here is capable of probing a broad spectral region from visible[45] to far-infrared energies[46]. First pump-probe spectroscopy data collected with this apparatus revealed ultrafast optical modulation of the infrared plasmonic response of graphene. This finding is a precondition for ultrafast graphene-based plasmonic devices. In terms of its efficiency, optical control of plasmons in graphene is on par with conventional electrostatic gating. Remarkably, the pulse energies needed to modify the infrared plasmonic response are two orders of magnitude smaller than what is typically necessary for comparable ultrafast switching times in metal-based plasmonic structures at NIR frequencies.[47] Notably, ultrafast plasmonic tuning in graphene can be readily accomplished using moderately priced fiber lasers, as first demonstrated here. A relevant figure of merit in this context is $R_{mod}$, defined by the ratio of the probe intensity modulation depth (%) to the necessary pump fluence (mJ/cm$^2$). We estimated $R_{mod} \sim 100$ for graphene: an unprecedentedly high value that exceeds $R_{mod} \sim 6$ for ultrafast schemes in metal structures[48] or $R_{mod} \sim 40$ for much slower (~ 100 ps) Si-based devices.[49] We remark that this value has been achieved close to the strong 1125 cm$^{-1}$ SiO$_2$ surface phonon where the plasmon-phonon-polariton behaves predominantly phonon-like with long polariton lifetimes.[26] Real-space plasmon propagation potentially relevant for plasmonic devices has been observed at lower frequencies[5,6] where still $R_{mod}$ ~ 25. Significant improvement could be expected from higher pump pulse energies and more efficient NIR light concentration using resonant tips.[50] Optimized tips would be beneficial since in our study an area ~$10^6$ times larger than the probed area under the tip was illuminated and the total amount of NIR field enhancement is unclear (see SI). Nevertheless, this first infrared pump-probe experiment beyond the diffraction limit demonstrates the capability of this technique and paves the way to the exploration of a wide range of problems in condensed matter physics, biology and chemistry.

## ASSOCIATED CONTENT

**Supporting Information.** Supporting experimental data and theory details. This material is available free of charge via the Internet at http://pubs.acs.org.

## AUTHOR INFORMATION

**Corresponding Author**

dbasov@physics.ucsd.edu (D.N.B)

**Notes**

All authors declare no competing financial interests.

## ACKNOWLEDGMENT

M.W. acknowledges financial support by the Alexander von Humboldt foundation. Research at UCSD is supported by ONR. The development of near field instrumentation at UCSD is supported by DOE-BES. A. S. R. and A. H. C. N. acknowledge DOE grant DE-FG02-08ER46512, ONR grant MURI N00014-09-1-1063, and the NRF-CRP award R-144-000-295-281.

Table 1. Summary of the modeling parameters for the Drude model of plasmonic response.

| | equilibrium weak MIR excitation | non-equilibrium strong MIR excitation | | | | |
|---|---|---|---|---|---|---|
| | | | + gating | + NIR excitation with power … | | |
| | | | | 2 mW | 5 mW | 10 mW |
| **Drude weight $D$** | $D_0$[a] | $1.10 D_0$ | $1.46 D_0$ | $1.21 D_0$ | $1.33 D_0$ | $1.46 D_0$ |
| $T$ (K) | 300 | 1500 | | 1700 | 1900 | 2100 |
| $\gamma$ (cm$^{-1}$) | 100 | 300 | | | | |

[a] $D_0 = 24.3|\mu_0|\sigma_0/\hbar$ with universal conductivity $\sigma_0 = e^2/(4\hbar)$ and $|\mu_0| = 1500$ cm$^{-1}$.

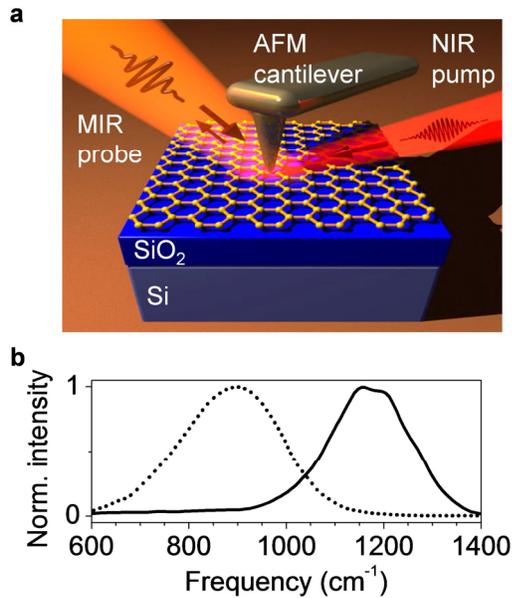

**Figure 1.** (a) Sketch of the NIR pump MIR probe near-field spectroscopy experiment on exfoliated graphene. (b) MIR spectroscopy data were acquired with probe pulses forming these two overlapping, broad-band spectra.

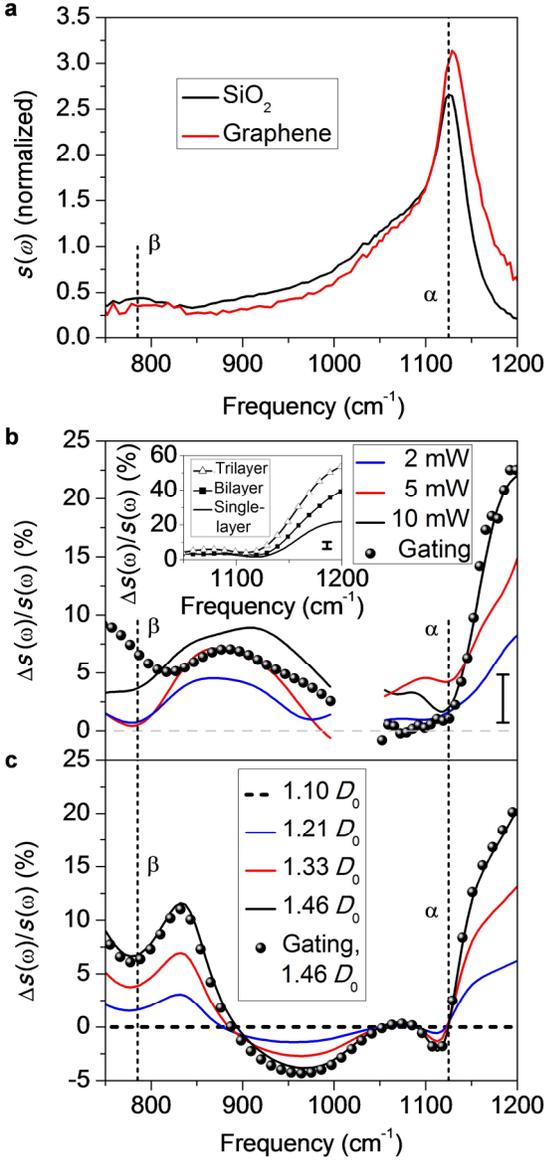

**Figure 2.** Graphene near-field spectra with and without NIR photoexcitation. (a) Near-field amplitude MIR spectra $s(\omega)$ (without NIR excitation) normalized to Si, for 300 nm $SiO_2$ (black curve) and for single-layer graphene on $SiO_2$ (red curve, probed with attenuated MIR intensity). The $SiO_2$ phonon modes at $\omega_\alpha = 1125$ cm$^{-1}$ and $\omega_\beta = 785$ cm$^{-1}$ are labeled as $\alpha$ and $\beta$. (b) NIR pump-induced, spectral changes in the near-field amplitude $\Delta s(\omega)/s(\omega)$ of graphene at zero time delay for varying NIR average powers. The effect of electrostatic gating (without NIR pumping)

is shown for comparison. The inset of panel (b) displays $\Delta s(\omega)/s(\omega)$ for different graphene films at 10 mW pump power. (c) Theoretical effect of increasing the intraband Drude weight $D$ above the initial non-equilibrium value $1.10D_0$ that serves as reference without NIR pump in intense MIR probe fields. While NIR pumping effects were modeled by an increase in carrier temperature, gating effects were simulated by adjusting the Fermi energy.

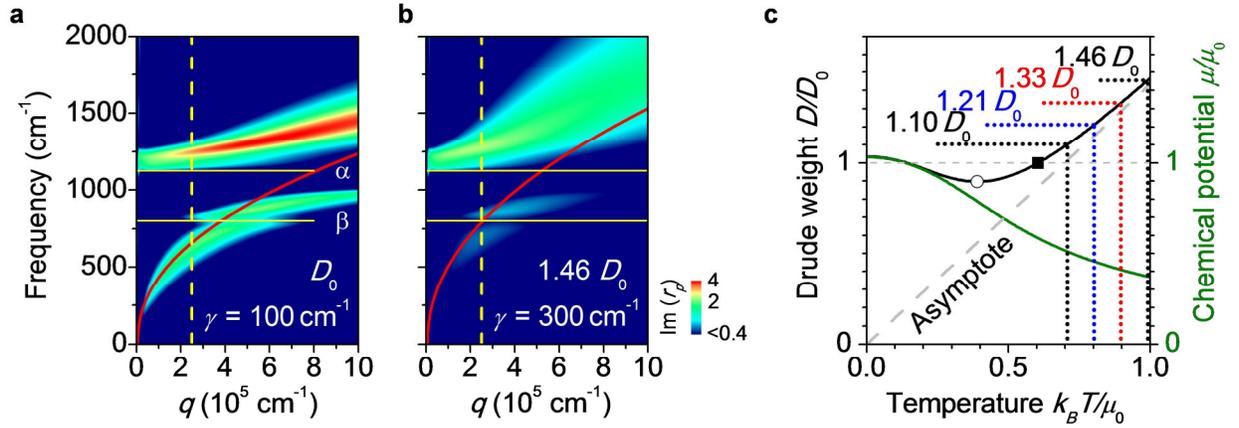

**Figure 3.** Calculated plasmon-phonon mode dispersion and Drude weight. (a) Graphene plasmon SiO$_2$ phonon mode dispersion evaluated as the imaginary part of the reflection coefficient $r_p(q,\omega) = 1 - \left(\kappa(\omega) + 2\pi i \sigma \frac{q}{\omega}\right)^{-1}$. The vertical dashed line marks the probe wavevector $q_{\text{tip}} \sim 1/a$ ($a = 40$ nm), the horizontal lines denote the $\alpha$ and $\beta$ SiO$_2$ phonons, and the red curve represents the graphene plasmon without phonon interaction. The calculated dispersion curves for the case of phonon interaction are given for the indicated Drude weight $D$ and scattering rate $\gamma$, corresponding to attenuated MIR intensity. (b) Full MIR probe intensity and synchronous 10 mW NIR pumping, characterized by an increase in $D$ and $\gamma$. (c) Theoretical, non-monotonic temperature dependence of $D$ and $\mu$ normalized to their $T = 300$ K or equilibrium values. The values of *D* obtained by modeling the experimental NIR pump induced spectral changes of Figure 2b are marked with the corresponding carrier temperature *T*, showing that *T* and *D* increase for NIR pumping. The minimum in *D* (open dot) is marked together with the position where the equilibrium value $D = D_0$ is recovered (black square).

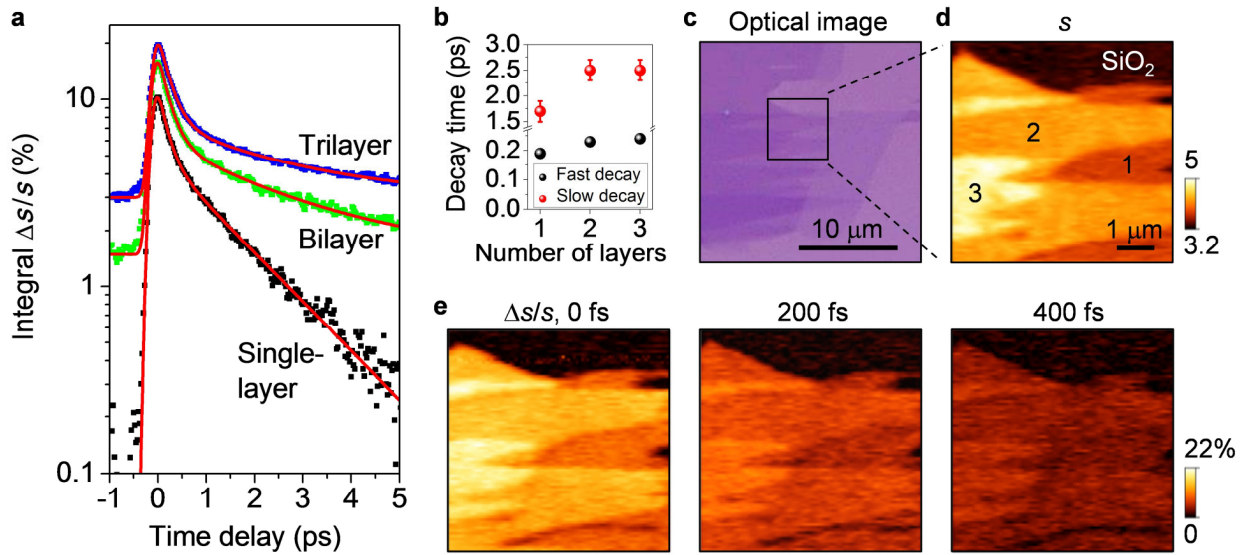

**Figure 4.** Nanoscale pump-probe dynamics and imaging. (a) Time-resolved, spectrally integrated changes Δ*s*/*s* in the near-field amplitude for 1-, 2- and 3-layer graphene. Probe pulses were tuned to cover the NIR pump-induced change at 1200 cm$^{-1}$ and 10 mW pump pulses were used for all panels. Traces are offset for clarity and biexponential fits are shown as red curves. (b) Time constants for the fast and slow decay components obtained from the fits in panel (a). (c) Optical image of the few-layer sample. (d) Scattering amplitude *s* showing 1-3 graphene layers (without NIR excitation). (e) Spatially resolved Δ*s*/*s* with NIR excitation for different pump-probe time delays.

**Supporting Information**

**Ultrafast and Nanoscale Plasmonic Phenomena in Exfoliated Graphene Revealed by Infrared Pump-Probe Nanoscopy**

*Martin Wagner[†], Zhe Fei[†], Alexander S. McLeod[†], Aleksandr S. Rodin[†,I], Wenzhong Bao[‡,$], Eric G. Iwinski[†], Zeng Zhao[$], Michael Goldflam[†], Mengkun Liu[†], Gerardo Dominguez[∥,⊥], Mark Thiemens[⊥], Michael M. Fogler[†], Antonio H. Castro Neto[□,I], Chun Ning Lau[$], Sergiu Amarie[▪], Fritz Keilmann[=], and D. N. Basov*[†]*

[†] Department of Physics, University of California, San Diego, La Jolla, California 92093, USA
[I] Department of Physics, Boston University, 590 Commonwealth Avenue, Boston MA 02215, USA
[‡] Materials Research Science and Engineering Center, University of Maryland, College Park, Maryland 20742, USA
[$] Department of Physics and Astronomy, University of California, Riverside, California 92521, USA
[∥] Department of Physics, California State University, San Marcos, San Marcos, California, 92096, USA
[⊥] Department of Chemistry and Biochemistry, University of California, San Diego, La Jolla, California 92093, USA
[□] Graphene Research Centre and Department of Physics, National University of Singapore, 117542, Singapore
[▪] Neaspec GmbH, Bunsenstr. 5, 82152 Martinsried, München, Germany
[=] Ludwig-Maximilians-Universität and Center for Nanoscience, 80539 München, Germany
* Corresponding author: dbasov@physics.ucsd.edu (D.N.B)

## 1. Experimental details

A schematics of the ultrafast setup based on a commercial s-SNOM system (NeaSNOM, Neaspec GmbH) is presented in Figure S1. Mid-infrared (MIR) light is generated by difference-frequency mixing in a 2 mm thick, z-cut GaSe crystal. To this end, 100 fs near-infrared (NIR) pulses of a 40 MHz Er-doped fiber laser ($\lambda = 1.56$ µm) are mixed with synchronized supercontinuum (SCIR) pulses at $\lambda = 1.8$ µm (Toptica FemtoFiber pro IR and SCIR).[1,2] The emitted MIR pulses have a spectral width of 200 cm$^{-1}$ full width at half maximum (FWHM). They are continuously tunable in the 650 - 2400 cm$^{-1}$ region and provide an average MIR power of up to 0.5-1.0 mW. These pulses are split in a Michelson interferometer comprising a scanning mirror in one arm and the scattering atomic force microscope (AFM) tip in the focus of a parabolic mirror in the other arm. The AFM tip is positioned above the sample, in this case exfoliated graphene on $SiO_2$. Backscattered light from the tip sample interaction region interferes with the reflected reference light at a mercury-cadmium-telluride (MCT) detector. An interferogram is recorded when moving the scanning mirror and subsequent Fourier transformation provides spectral information. Near-field signals $s$ are obtained by standard demodulation of the tip oscillation at the 2$^{nd}$ harmonic. Commercial PtIr coated tips with nominal tip radii of $a$ = 25 nm are used in tapping mode at frequencies around 250 kHz with an amplitude of 60 nm. The NIR beam can also serve as a pump. Since it originates from the same laser feeding the DFG crystal it is perfectly synchronized with the MIR probe beam. Relative timing between pump and probe is controlled with a mechanical delay stage in the NIR beam path. With an elliptical spot of ~20 x 40 µm$^2$ (FWHM) the pump beam is focused onto the AFM tip in $p$-polarization (same as MIR beam). All measurements are performed under ambient conditions. The lattice temperature of the studied specimens remains below ~330 K under NIR pumping, as estimated from the thermally driven metal-to-insulator transition in $VO_2$ films under similar conditions.[3]

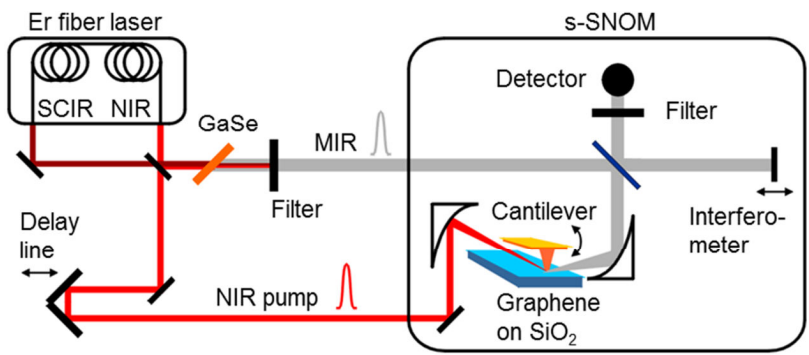

**Figure S1.** Setup for ultrafast nanoscale spectroscopy and imaging. Mid-infrared (MIR) probe pulses obtained from difference-frequency mixing of near-infrared (NIR) and supercontinuum (SCIR) pulses in GaSe are focused on the tip of an s-SNOM together with NIR pump pulses. The tip-scattered MIR light containing near-field information is analyzed with a Michelson interferometer.

Fitting of the pump-probe temporal response in Figure 4a of the main text reveals an MIR pulse duration of (200 ± 10) fs (FWHM) in accord with a Fourier-transform-limit of 75 fs broadened by dispersion of the optical components of our system. From the pulse length we estimated an MIR peak field strength of 30 kV/cm. An additional tip-induced enhancement of the electric field by a factor of ~30 is suggested by a Comsol simulation, giving rise to a total, in-plane peak field strength of the order of $10^2$ kV/cm. We note that these fields are remarkably intense given the 40 MHz repetition rate and <1 mW average power delivered by our laser system.

Attenuation of the MIR probe pulse intensity by a factor of ~40 was achieved primarily by pulse elongation in a dispersive, but transmissive NaCl slab. This method ensures minimal loss of the near-field signal compared to an alternative approach involving a reduction of the average power. However, temporally stretching the MIR pulses up to ~5 ps is incompatible with the high time resolution that is required to monitor the sub-picosecond carrier dynamics in graphene. Therefore, pulse elongation is avoided for the NIR pump MIR probe experiment described in the main text. Instead, we use the unattenuated probe beam though it modifies the response of graphene.

We study exfoliated graphene samples in the pump-probe experiment in contrast to the commonly used large-scale epitaxial and chemical-vapor deposited graphene that usually exhibits less favorable electronic properties, but is necessary for diffraction-limited experiments. The single- and few-layer graphene samples on Si substrate with 300 nm $SiO_2$ layer are derived from exfoliation according to standard procedures. Raman spectra show unintentional hole-doping and the single-layer samples have a hole density of $|n|$ = (2.6 ± 1) × $10^{12}$ cm$^{-2}$ corresponding to a chemical potential of $\mu_0$ = -(1500 ± 300) cm$^{-1}$.

Electrostatic gating uses the Si substrate as a back-gate while the tip and graphene are grounded. For the employed gate voltage $V_g$ = -70V we deduce an induced carrier density of $|\Delta n|$ = 5 x $10^{12}$ cm$^{-2}$ from the common capacitor model.[4] Note that this value can differ significantly from the actual carrier density under ambient conditions.[5]

For normalization of the near-field spectra, a sample of a 300 nm SiO$_2$ film on Si with trenches etched down to the Si substrate serves as a reference. To this end near-field spectra of SiO$_2$ and adjacent Si are measured and the normalized amplitude $s^{SiO2}(\omega)/s^{Si}(\omega)$ is used to reference graphene spectra $s^{graphene}(\omega)/s^{SiO2}(\omega)$ (previously normalized to adjacent SiO$_2$) to Si.

## 2. Frequency-resolved acquisition mode and spectrally integrated acquisition mode

The time-resolved s-SNOM signal is acquired in two different modes of operation. One method entails measurements of the scattering amplitude spectra $s(\omega)$. We refer to this method as the frequency resolved data acquisition mode. These spectra are collected at varying time delays $\Delta t$ between MIR probe and NIR pump pulses. We introduce the following notation: $s(\omega)_{\Delta t=-5ps}$ denotes the spectrum taken with the probe pulse preceding the pump pulse by 5 ps; $s(\omega)_{\Delta t=0}$ is the spectrum measured at zero temporal overlap, etc. In the former case the probed spectrum is unaffected by the subsequent pump pulse. Pump-induced, differential changes $\Delta s(\omega) = s(\omega)_{\Delta t=0} - s(\omega)_{\Delta t=-5ps}$ are most informative for our analysis.

A complementary method is the frequency integrated data acquisition mode that is achieved by blocking the reference arm of the Michelson interferometer. This method yields significant improvement of the signal-to-noise ratio at the expense of spectral information. This compromise is still advantageous for tasks requiring fast imaging. A mechanical chopper modulates the NIR beam. A lock-in amplifier referenced to the cantilever oscillation detects the 2$^{nd}$ harmonic $s_2$ while an additional lock-in amplifier synchronized to the pump beam modulation detects the differential change $\Delta s$ within $s_2$. A potential issue arising from blocking the interferometer arm is that the detected s-SNOM signal is no longer background-free. This signal contains the field directly scattered by the tip (subject to near-field interaction) *multiplied* with other, i.e. background fields.[6] The latter are insensitive to the pump-probe delay as witnessed by the absence of pump-induced changes in the SiO$_2$ substrate in Figure 4 of the main text. Consequently, the near-field relaxation dynamics of Figure 4 is not affected by the spurious background.

## 3. Nonlinear response of graphene in strong MIR probe fields

Figure S2a compares $s(\omega)$ spectra of graphene on SiO$_2$ probed with full MIR intensity (red curve) and after attenuating the probe beam by a factor of ~40 (blue curve). Both spectra are dominated by the SiO$_2$ surface phonon resonances (black curve with α and β marking the phonon positions). Surprisingly, only in the case of an attenuated probe beam were we able to reproduce the data obtained with a continuous wave (cw) quantum cascade laser (Daylight solutions, blue triangles in Figure S2a).

In order to understand the underlying physics we recall that the two-dimensional electron gas in graphene is prone to at least two forms of photo-induced effects: interband photo-generation and intraband heating of carriers.[7-12] The former process is negligible here because the relevant MIR probe frequencies (probe spectra in Figure 1b of the main text) correspond to at most half of the lowest energy of interband transition at $2|\mu_0|$. Here $\mu_0 = -1500$ cm$^{-1}$ is the chemical potential at $T = 300$ K for our specimens. Therefore the creation of a hot carrier distribution[12,13] is the dominant effect in the intense MIR fields. We modeled the impact of increased carrier temperature ($T = 1500$ K) in Figure S2b by introducing an enhanced electron damping[14] $\gamma = 300$ cm$^{-1}$ in the intraband Drude model of plasmonic response. Details of the modeling procedure are described in Section 4. While the corresponding Drude weight of 1.10 $D_0$ exceeds the equilibrium value $D_0$ by 10%, the scattering rate reveals a much stronger, three-fold increase compared to the equilibrium value. The net effect are significantly broadened plasmon-phonon resonances that account for our observations in Figure S2.

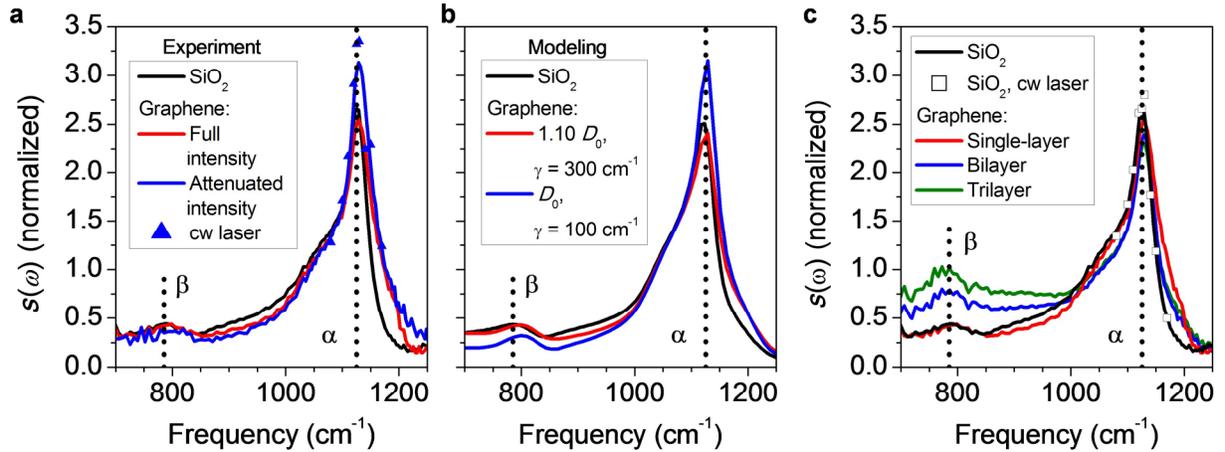

**Figure S2.** Impact of the probe-beam intensity on the spectral characteristics of the graphene-SiO$_2$ interface. (a) Normalized near-field amplitude spectra $s(\omega)$ for 300 nm SiO$_2$ on Si substrate (black curve) and for single-layer graphene on top of SiO$_2$/Si (red curve). These data were obtained at full MIR intensity. A discrepancy between the red spectrum and the cw data (triangles) is immediately apparent. Once the MIR probe beam is attenuated by a factor of ~40 (blue curve) we restored excellent agreement with the cw results. (b) The calculated spectra reproduce the experimental data in panel (a). The graphene response probed at full MIR intensity is reproduced assuming an enhanced Drude weight (measured in units of $D_0$ as defined in the text) and scattering rate $\gamma$. The increase of both the Drude weight and the scattering rate is consistent with the notion of the rise of electronic temperature of the sample triggered by the intense electric field of the MIR beam. Details of the calculations are described in Section 4. (c), Spectra of bilayer (blue line) and trilayer graphene (green line) are compared with those of single-layer graphene (red line) and SiO$_2$ (black line), all obtained at full MIR intensity. The SiO$_2$ response is identical when probed with either a cw (symbols) or the intense pulsed MIR laser (black line).

## 4. Modeling

Our theoretical simulations of the near-field signal follow the procedure introduced in refs. 15, 16. The underlying physical picture is that MIR illumination polarizes the AFM tip, which then emits evanescent waves with momenta $q_{tip} \sim 1/a$. These fields are modified in the presence of a conductive or polar medium below the tip. In turn, this changes the polarization of the tip and the backscattered radiation arriving at the detector. In principle, the precise amplitude of the near-field signal depends not only on the tip radius $a$ but also on the overall shape of the tip, in our case modeled as an elongated metallic spheroid. This latter approach gives a reasonable compromise between the amount of details and the associated computational cost.[16] Since the incoming light is predominantly $p$-polarized, the tip-sample interaction is described by the reflection coefficient $r_p(q,\omega)$.[15] The input parameters for computing $r_p$ are the in-plane conductivity of graphene and the optical constants of Si and $SiO_2$. The latter were taken from far-field literature data.[17] To model the conductivity of graphene $\sigma(T) = \sigma_{intra}(T) + \sigma_{inter}(T)$ we use the following equations for our effective temperature model[18-20]:

$$\sigma_{intra}(T) = \frac{i}{\pi}\frac{D}{\omega + i\gamma}, \quad D = 8\frac{\sigma_0}{\hbar}k_B T \ln\left[2\cosh\left(\frac{\mu}{2k_B T}\right)\right], \tag{S1}$$

$$\sigma_{inter}(T) = \sigma_0 H\left(\frac{\omega}{2}\right) + \frac{4i\omega}{\pi}\sigma_0 \int_0^\infty dx \, \frac{H(x)-H\left(\frac{\omega}{2}\right)}{\omega^2 - 4x^2}, \tag{S2}$$

$$H(x) = \frac{\sinh\left(\frac{\hbar x}{k_B T}\right)}{\cosh\left(\frac{E_F}{k_B T}\right) + \cosh\left(\frac{\hbar x}{k_B T}\right)}. \tag{S3}$$

Here $\sigma_0 = e^2/(4\hbar)$ is the universal conductivity and $\gamma$ is the scattering rate. Our near-field spectra are dominated by the intraband response $\sigma_{intra}$ (eq S1). In order to apply eq S1 to the analysis of our data, we need to know how the chemical potential $\mu$ varies with temperature $T$. Our principal assumption is that the carrier concentration, i.e. the electron density in the conduction band *minus* the hole density in the valence band, remains fixed. Only carrier diffusion could change the carrier densities. For the optimistic estimate of a transport mobility $\mu_{tr} \sim 10.000$ cm$^2$/Vs, the momentum relaxation time is given by $\tau = E_F \mu_{tr}/(ev_F^2) \approx 200$ fs (with Fermi velocity $v_F$). A mean-free path of $\tau v_F \approx 200$ nm follows. Thus, diffusion across the photoexcitation spot of 20 μm diameter is beyond the ~ps time scale of interest in our case. On the other hand, an increased carrier density spatially confined directly under the tip and originating from the tip's field-enhancement would be obscured by carrier diffusion away from the ~20 nm probe spot. This occurs on much faster time scales than our 200 fs time resolution, justifying the assumption of constant carrier density and preventing an estimation of the field enhancement factor for the pump beam.

The constraint of the fixed electron concentration $n$ translates into:

$$\int_0^\infty d\epsilon \, \frac{2\epsilon}{\pi v_F^2 \hbar^2} [f(\epsilon, \mu) - f(\epsilon, -\mu)] = n, \tag{S4}$$

where $2\epsilon/(\pi v_F^2 \hbar^2)$ is the density of states and $f(\epsilon, \mu) = [1 + \exp((\epsilon - \mu)/k_B T)]^{-1}$ is the Fermi-Dirac distribution. Integration over the conduction and valence band and application of the relation $\pi v_F^2 \hbar^2 n = \mu(0)$ leads us to the transcendental equation for $\mu(T)$:

$$\mu^2(T) + k_B^2 T^2 \left[\frac{\pi^2}{3} + 4\mathrm{Li}_2\left(e^{-\mu/k_B T}\right)\right] = \mu^2(0), \tag{S5}$$

where $\mathrm{Li}_2(z)$ is the dilogarithm function. Approximate analytical solutions exist in the limits of low and high temperature:

$$\frac{\mu(T)}{\mu(0)} = \begin{cases} 1 - \frac{\pi^2}{6} \frac{k_B^2 T^2}{\mu^2(0)}, & T \ll \mu(0), \\ \frac{1}{4 \log 2} \frac{\mu(0)}{k_B T}, & T \gg \mu(0). \end{cases} \tag{S6}$$

At intermediate $T$, a numerical solution can be easily obtained. In Figure 3c in the main text we reference $\mu(T)$ to its equilibrium or $T = 300$ K value $\mu_0$ instead of using the (experimentally inaccessible) zero-temperature $\mu(0)$ value. The ratio $\mu(T)/\mu_0$ is a decreasing function of the dimensionless combination $k_B T/\mu_0$. This behavior is typical of systems where the electron density of states increases with the (absolute value of) the Fermi energy $\mu(0)$, e.g., the usual 3D Fermi gas.

For $T \to 0$, eq S1 reduces to $D = \left(\frac{e^2}{\hbar^2}\right) \mu \propto \sqrt{|n|}$ (with carrier density $n$)[21] which can be represented in a more familiar form, $D = \pi |n| e^2/m^*$, valid for a degenerate electron gas with the cyclotron mass $m^* = \frac{\hbar k_F}{v_F} \propto \sqrt{|n|}$. At high temperatures, $k_B T \gg \mu$, eq S1 implies $D \propto T$, which can be understood based on another classical formula $D = \left(\frac{\pi e^2}{2 k_B T}\right)\left(n_e \langle v_e^2 \rangle + n_h \langle v_h^2 \rangle\right)$ suitable for a non-degenerate plasma of electrons (holes) with density $n_e$ ($n_h$). In graphene the thermal velocity averages $\langle v_e^2 \rangle$ and $\langle v_h^2 \rangle$ are replaced by $v_F^2 = $ const, while the thermally excited particle densities scale as $n_e \simeq n_h \propto T^2$, leading to the overall linear dependence of $D(T)$. Therefore, at high photoexcitation where this regime is realized, the Drude weight is expected to increase with absorbed power. Combining eq S1 and S5 we obtain the full $D(T)$ dependence. $D(T)$ is non-monotonic, with a minimum at $T_{min} = 0.39 \, |\mu_0|/k_B$ (open dot in Figure 3c of the main text). At large temperatures, $D(T)$ follows the linear law (asymptote in Figure 3c) characteristic of the charge-neutral graphene[18]: $D(T) \simeq (2 \ln(2 e^2)/\hbar) k_B T$.

The equilibrium temperature value of the Drude weight is restored at $T = 0.60\,|\mu_0|/k_B$ (black square in Figure 3c of the main text), corresponding to 1270 K for our case of $|\mu_0| = 1500$ cm$^{-1}$. If the probe pulse does heat the sample above this temperature, only an increase in $D$ would be observed. This scenario seems to be the case realized in our experiment. The non-monotonic behavior of $D$ suggests a region with $T < T_{min}$ where $D$ and hence the near-field signal are expected to decline with increasing carrier temperature, a temperature range apparently not accessible by our experimental conditions.

Note that our modeling contains the following simplifications. It disregards all dynamics, i.e. ultrafast thermalization and cooling are not accounted for, that are assumed to occur on time scales equal to the probe pulse duration. Many-body effects[22] are not included and the theory was derived for $q = 0$. Additionally, MIR interband inversion[7, 23] is neglected along with possible MIR multiphoton interband transitions. These shortcomings notwithstanding, our model qualitatively captures the key features of our experiment.

## 5. Modeling the scattering near-field amplitude and differential spectra

We started by modeling spectra of SiO$_2$ normalized to Si using the model outlined in Section 4. We treated the tip radius and cantilever tapping amplitude as adjustable parameters. The results of this modeling are displayed in Figure S2b. The best value for the tapping amplitude in order to reproduce the experimental data was 60 nm and for the tip radius $a = 40$ nm; these agree well with nominal values. Then both parameters were kept constant when simulating the graphene spectra without NIR illumination. For $|\mu_0| = 1500$ cm$^{-1}$ we varied $T$ and $\gamma(T)$ and were able to adequately reproduce experimental data in Figure S2b. We report the modeling parameters directly related to the key findings of our work in Table 1 in the main text.

In Figure S3a we justify our choice of parameter values for the case of MIR probing at full intensity, still without NIR excitation. Model spectra for different $T$ and $\gamma$ values are compared to the measured ones, with the best overall agreement given for $T = 1500$ K and $\gamma = 300$ cm$^{-1}$. Figure S3a also demonstrates the impact of varying $T$ and $\gamma$. The enhanced $T$ slightly increases the spectral weight of both the α and β modes. The increase of $\gamma$ broadens and reduces the α mode and enhances the β mode.

In Figure S3b we assessed the impact of the interband conductivity for our model with $T = 1500$ K and $\gamma = 300$ cm$^{-1}$. A concerted action of both intra- and interband contributions according to eqs S1-S3 produces the black trace that differs by less than 10% from a simplified intraband Drude model based on eq S1 alone (red line). This latter result highlights the dominance of intraband conductivity for our experimental conditions.

Turning now to NIR excitation we set again $|\mu_0| = 1500$ cm$^{-1}$ for all powers. Throughout our simulation we keep the damping $\gamma$ constant, although one could expect some increase with excitation density and electronic temperature.[14] The inset of Figure S3b demonstrates the rationale behind our choice of the damping parameter. Here we plot pump-induced changes in the near field spectra corresponding to 10 mW pump power for some $\gamma$ values. Scattering rates different from $\gamma = 300$ cm$^{-1}$ (dotted line), the value that best reproduces the spectra without NIR excitation, produce spectra strongly departing from experimental data. We speculate that damping in the presence of MIR probe pulses alone is already significant, so that a further increase is either not present or not distinguishable here.

The real ($\sigma_1$) and imaginary parts ($\sigma_2$) of the conductivity calculated using modeling parameters in Table 1 in the main text are presented in Figure S4a. Throughout the relevant spectral region of our experiment (750 - 1200 cm$^{-1}$; see magnified inset) we observe a positive photoconductivity $\Delta\sigma_1$ for MIR and NIR excitation in accord with earlier observations in the THz range.[24,25] However, recent THz experiments report the opposite behavior: a negative photoconductivity.[8,10,26,27] We obtain the latter situation for frequencies below 150 cm$^{-1}$ (115 cm$^{-1}$) for MIR (NIR) photoexcitation. The different signs of the photoconductivity in literature were attributed to differences in scattering rates.[26]

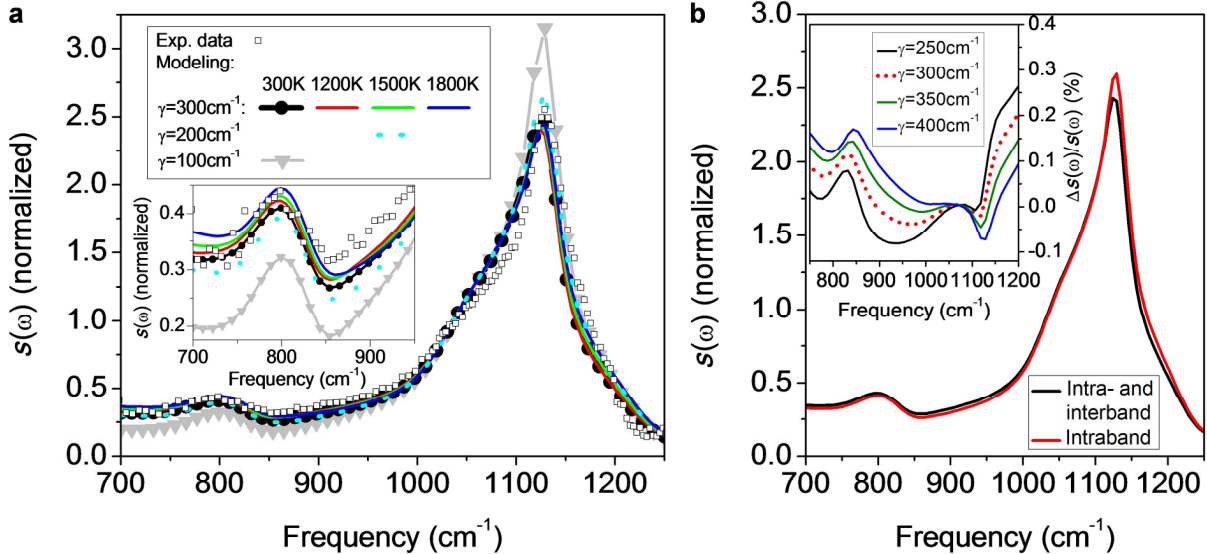

**Figure S3.** Modeling of graphene spectra. (a) Experimental data for the near-field scattering amplitude $s(\omega)$ of graphene obtained at full MIR probe intensity (open squares). Modeling results for different temperature and damping values are presented with color lines. The low frequency region is magnified in the inset. Best agreement with experimental data is achieved for $T = 1500$ K and $\gamma = 300$ cm$^{-1}$. (b) Near-field response of graphene on SiO$_2$ substrate calculated using a model of intraband only response (red curve) and complete intra-/inter-band conductivity (black curve). In the inset we show model simulations of the spectral changes $\Delta s(\omega)/s(\omega)$ produced by 10 mW NIR excitation. These results were obtained for a Drude weight of 1.46 $D_0$ for different values of the scattering rate. Best agreement with overall experimental trends is found for $\gamma = 300$ cm$^{-1}$ (dotted line).

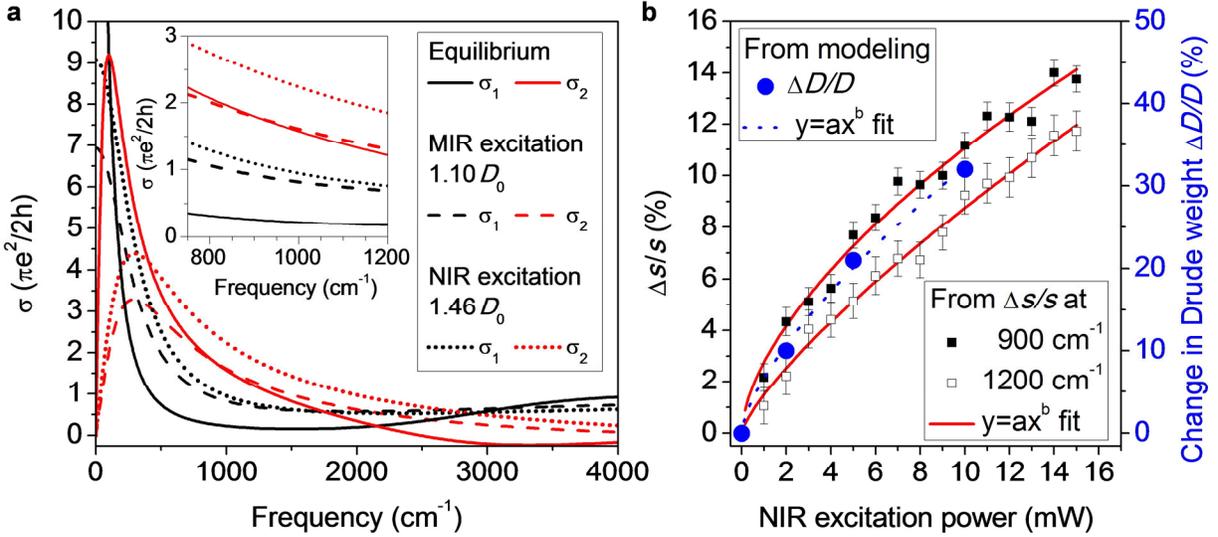

**Figure S4.** Optical conductivity at various excitation conditions. (a) Modeled conductivities for equilibrium conditions ($T = 300$ K, $\gamma = 100$ cm$^{-1}$) (full line), non-equilibrium MIR photoexcitation ($T = 1500$ K, $\gamma = 300$ cm$^{-1}$, dashed) and 10 mW NIR photoexcitation ($T = 2100$ K, $\gamma = 300$ cm$^{-1}$, dotted). In the frequency region relevant for our experiments (see inset) the real part $\sigma_1$ increases under either MIR or NIR excitation. (b) Spectrally integrated changes in the near-field amplitude $\Delta s/s$ as a function of NIR pump power. Data were taken at zero time delay between pump and probe with the probe beam centered at 900 cm$^{-1}$ (black squares) and 1200 cm$^{-1}$ (open squares). Experimental data and $\Delta D/D$ (blue dots) from modeling show sublinear scaling with excitation power as discussed in the text.

## 6. Power dependence of NIR pump-induced changes

In Figure 2b of the main text, we plot spectra of $\Delta s(\omega)/s(\omega)$ exhibiting sublinear scaling with the applied pump power. For a more quantitative analysis, integral changes of $\Delta s/s$ are presented in Figure S4b. These data were obtained by means of the spectrally integrated method 2 (Section 2) with the probe beam centered at 900 cm$^{-1}$ (black squares) and 1200 cm$^{-1}$ (open squares) at zero temporal overlap $\Delta t = 0$. Fitting a power law $y = ax^b$ we obtain exponents $b_{900\text{cm}^{-1}} = 0.60 \pm 0.03$ and $b_{1200\text{cm}^{-1}} = 0.77 \pm 0.03$, revealing a sublinear dependence on excitation power.

At the maximum fluence of ~1 GW/cm$^2$, saturation of the pump light absorption has not set in yet.[28] In fact, the sublinear power behavior is also present in the NIR pump induced Drude weight changes $\Delta D/D$ obtained in our model (blue dots). The corresponding power law exponent reads $b = 0.69 \pm 0.06$, close to the above values. Thus, the power-dependence of the intraband conductivity is directly translated to the power-dependence of the near-field observable $\Delta s/s$. Note that a similar relationship between conductivity and NIR excited electron-hole pair density has been found recently in another study.[8]